\documentclass[11pt]{article}
\usepackage{amsmath,amsfonts, amssymb,graphicx,geometry,authblk,color,soul,comment,hyperref} 
\newcommand{\tr}{\text{tr}}

\title{Artificial sunflower: \\ Light-induced deformation of photoactive shells}
\author{Sathvik Sanagala and Kaushik Bhattacharya}
\affil{California Institute of Technology}

\begin{document}
\maketitle

\begin{abstract}

Photomechanically active materials undergo reversible deformation on illumination, making them ideal for remote, tether-free actuation.   Much of the work on these materials has focused on one-dimensional structures, such as strips.  In this paper, we explore photomechanically active two-dimensional structures such as  sheets and shells.  When illuminated, such structures undergo spontaneous bending due to the limited penetration of light.  However, the geometry of the shell constrains possible deformation modes: changes in Gauss curvature lead to in-plane stretching, against which shells are very stiff.  Therefore, there is a complex coupling between the photomechanical actuation and the mechanical behavior of a shell.  We develop and implement a novel approach to study photomechananically active shells. This method is a discrete shell model which captures the interplay between actuation, stretching, bending, and geometric changes.  Through a series of examples, we explore these complex interactions, demonstrating how one can design shells that deform to follow the source of illumination.

\end{abstract}

\section{Introduction}
Stimuli-responsive materials have received increased attention due to their ability to convert external fields into mechanical work, offering promising applications including artificial muscles \cite{Mirvakili_2018}, soft robotics \cite{Laschi_2016}, and actuators \cite{Hu_2019}. Within this class, photomechanically active materials are of particular interest due to their capacity to deform when exposed to light, allowing for remote, tether-free actuation \cite{white_photomechanical_2017}.  An important class of photomechanically active materials are liquid crystal elastomers, which are embedded with photoactive molecules such as azobenzene \cite{Yu_2003,Gelebart_2017,white_photomechanical_2017}.  

Liquid crystal elastomers are crosslinked polymer networks which incorporate rod-like liquid crystal molecules (mesogens) into their polymer chains \cite{warner_liquid_2003}.  These materials undergo an isotropic to nematic transformation accompanied by a large shape change.   When photoactive molecules are embedded into the polymer structure, liquid crystal elastomers undergo a light-induced deformation due to the illumination-controlled \textit{trans}-to-\textit{cis} isomerization of these molecules \cite{corbett_2006,white_photomechanical_2017,bai_2020}.  Azobenzene is a common choice for photomechanically functionalizing liquid crystal elastomers because it has a high absorption coefficient across various wavelengths with fast photo-chemistry  \cite{Kizilkan_2016}.   Lower wavelengths ($\sim$300-400nm) of ultraviolet light induces to \textit{trans}-to-\textit{cis} isomerization while visible illumination ($>$400nm) reverses the process, causing \textit{cis}-to-\textit{trans} isomerization.  These transitions couple with the nematic order of the liquid crystal elastomer, resulting in photomechanically-induced deformation.   The azobenzene molecules are linear in shape in the \textit{trans} state, and bent in shape in the \textit{cis} state.   Consequently, the {\it trans} state promotes the nematic order in the liquid crystal elastomer while the {\it cis} state reduces the nematic order.   Further, illumination with linearly polarized light selectively isomerizes the appropriately oriented azobenzene molecules, further promoting nematic director alignment.  This change in nematic order and director alignment is accompanied by spontaneous stretch.

When a sheet of photomechanically active material is illuminated, the light intensity decreases with depth as it absorbed, resulting in limited penetration.  When the penetration depth is small, the intensity is described by Beer's law 
\begin{equation}
    I = I_{0}e^{-\frac{x}{d}},
\end{equation}
where $I_{0}$ is the light intensity at the illuminated surface, $x$ is the position within the material (where $x = 0$ corresponds to the surface of illumination), and $d$ is the penetration depth \cite{Warner_2014}.  The photo-isomerization decays throughout the thickness, and consequently, the deformation decays as well, resulting in the sheet bending \cite{Yu_2003,Warner_2014, Korner_2020}.  When the penetration depth is large, there can be non-monotone dependence on illumination \cite{Warner_2014,Maghsoodi_2024}.  In this work, we assume shallow penetration and the Beer's limit.

The light-induced bending deformation leads to a non-local shape change, which in turn alters the illumination conditions across the deformed structure.  This phenomenon creates complex deformations of a photomechanical strip under illumination \cite{white_photomechanical_2017}.   Gelebart {\it et al.}  \cite{Gelebart_2017} exploited this behavior to generate light-induced, periodic flapping of a doubly clamped liquid crystal elastomer film, and proposed that such motion can be used for a remotely actuated walker.  Korner  {\it et al.} \cite{Korner_2020} developed a theoretical model which demonstrated how steady illumination can give rise to periodic motion.  However, these studies have focused on one-dimensional strips and beams.

In this paper, we study photomechanically active sheets and shells, or two-dimensional structures.  The behavior of such structures is expected to be more complex due to the inherent coupling between stretching and bending, as a consequence of Gauss's remarkable theorem \cite{Gauss_1828}.  In recent years, the thermally induced shape-morphing of sheets and shells has been the topic of intense interest \cite{modes_2010,modes_2011,Ware_2015,ambulo_2017,Aharoni_2018}. Applying heat to liquid crystal elastomer sheets with a patterned director gives rise to non-uniform spontaneous stretch, resulting in out-of-plane deformations.  In contrast, photomechanical actuation induces spontaneous bending, altering the Gauss curvature and, consequently, the stretch.  This coupling can give rise to instabilities, necessitating care when studying these deformations.

There are numerous approaches for studying the mechanics of plates and shells \cite{Audoly_2000,Reddy_2006,Timoshenko_1959}.  The F{\"o}ppl-von K{\'a}rm{\'a}n equations \cite{Foppl_1921,Karman_1907} are commonly used for small deformations and shallow shells, but the study of large deformation requires higher order non-linear differential equations.  This in turn requires higher order discretization, resulting in stiff numerical problems.  An alternate approach with origins in computer graphics \cite{Baraff_1998,Grinspun_2003} beings in the discretized setting, and yields in lower order model.  In recent years, a strong connection to differential geometry has risen, enabling the development of discrete models for shells \cite{Crane_2018,Chen_2018}.  We integrate this approach with the energetics of shells to develop an efficient yet physically meaningful computational method for studying photoactive shells.  We apply this method to a series of examples involving photomechanical sheets and shells to demonstrate the richness of phenomena associated with the interplay between photoactuation, bending, and stretching.  A noteworthy example is one where a shell follows the source of illumination, much like a sunflower follows the sun.

The paper is organized as follows.  We present our formulation in Section \ref{sec:theory} starting with the continuous formulation in Section \ref{sec:cont} and the discrete formulation in Section \ref{sec:disc}.  We present examples with isotropic flat sheets in Section \ref{sec:flatiso}, flat anisotropic sheets in Section \ref{sec:flataniso} and curved shells in \ref{sec:shell}.    We conclude in Section \ref{sec:conc} by recalling and discussing the main results.

\section{Theory} \label{sec:theory}

In this section, we present our formulation of a photomechanically active shell.  We first present the theory in the continuous setting in Section \ref{sec:cont}.  We then present the discrete shell formulation that we use for our simulations in Section \ref{sec:disc}.

\subsection{Continuous Formulation} \label{sec:cont}

\paragraph{Kinematics}

\begin{figure}
  \centering
  \includegraphics[width=4in]{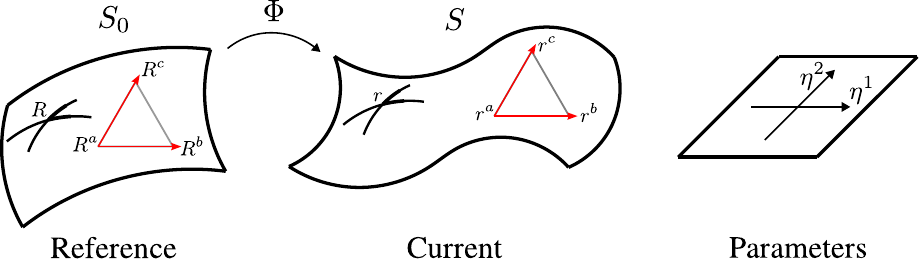}
  \caption{Kinematics. The reference shell is represented through its midsurface $S_0$ where the position of a particle $R$ is parameterized by parameters $\{\eta^1,\eta^2\}$. The deformation $\Phi$ maps $R$ on $S_{0}$ to the deformed position $r$ on the midsurface of the deformed shell $S$ .  The reference and current images of a discrete element is also shown with vertices $\{R^{a},R^{b},R^{c}\}$ and $\{r^{a},r^{b},r^{c}\}$ respectively.\label{fig:config}}
\end{figure}

We start with the Kirchhoff-Love assumption \cite{Reddy_2006}, which allows us to represent the shell volume in terms of the midsurface.  Therefore, we consider a shell to be a two-dimensional manifold in three-dimensional space.  We denote the reference domain occupied by the shell as $S_0\subset \mathbb{R}^3$ and current domain as $S \subset \mathbb{R}^3$, see Figure \ref{fig:config}. We label particles  $R\in S_0$ in the reference configuration, and $r \in S$ in the current configuration.  Therefore the deformation is the map $\Phi: S_0 \to S$ or $r= \Phi(R)$.

We introduce a parameterization  of the reference shell $R(\eta^{1},\eta^{2})$.  The deformation induces a parameterization  
\begin{equation} \label{eq:par}
r(\eta^{1},\eta^{2}) = \Phi (R(\eta^{1},\eta^{2}))
\end{equation}
of the current configuration.  We can now obtain a covariant bases $\{E_{\alpha}\}_{\alpha = 1}^{2}$ and $\{e_{\alpha}\}_{\alpha = 1}^{2}$ for the tangent spaces of the reference and current configurations respectively, as 
\begin{equation}
    E_{\alpha} = \frac{\partial R}{\partial \eta^{\alpha}}, \quad    e_{\alpha} = \frac{\partial r}{\partial \eta^{\alpha}}, \quad \alpha = 1,2.
\end{equation}
We denote the corresponding contravariant bases as $\{E^{\alpha}\}_{\alpha = 1}^{2}$ and $\{e^{\alpha}\}_{\alpha = 1}^{2}$ respectively ($E_{\alpha}\cdot E^{\beta} =  e_{\alpha}\cdot e^{\beta} = \delta_{\alpha}^{\beta}$).

We seek to find the stretch $g$ between the current and reference configurations, and the material curvature $h$.  These are defined as
\begin{equation} \label{eq:gh1}
    g = (\nabla_{R}r)^{T}(\nabla_{R}r) , \quad   
    h = -(\nabla_{R}r)^{T}(\nabla_{R}n) 
\end{equation}
respectively.  Above. $n = \left ( e_1 \times e_2\right )/|e_1 \times e_2|$ is the unit normal to $S$.
To obtain representations for these, we differentiate (\ref{eq:par}) with respect to $\eta^\alpha$ to conclude $e_{\alpha} = (\nabla_{R}r) E_{\alpha}$; it follows $(\nabla_{R} r)= e_{\alpha} \otimes E^{\alpha}$.  Similarly, $\partial n/\partial \eta^\alpha = (\nabla_{R}n) E_{\alpha}$; it follows $(\nabla_{R}n) =  \partial n/\partial \eta^\alpha  \otimes  E^{\alpha}$.  Therefore,
    \begin{equation} \label{eq:gh}
    g = \sum_{\alpha,\beta}a_{\alpha\beta}(E^{\alpha} \otimes E^{\beta}), \quad 
    h = -(\nabla_{R}r)^{T}(\nabla_{R}n) = \sum_{\alpha,\beta}b_{\alpha\beta} (E^{\alpha} \otimes E^{\beta}).
\end{equation}
where $ a_{\alpha\beta} = e_{\alpha} \cdot e_{\beta}$ and $b_{\alpha\beta} = -e_{\alpha}\cdot (\partial n/\partial \eta^{\beta})$ are the first and second fundamental forms of the current configuration.
Note that when there is no deformation or $\Phi = id$, then $g$ is identity and $h$ is the reference curvature.

We have used a global parameterization  for our definitions, but this is not necessary: one only needs a local parameterization.

\paragraph{Energy and Spontaneous Curvature}
We consider a Koiter type energy \cite{Koiter_1966}, allowing us to compute the total energy as the sum of a stretching and bending energy: 
\begin{equation}
    \mathcal{E} = \int_{S_0} \left(\frac{t}{4}W_{s} + \frac{t^{3}}{12}W_{b}\right)dA,
\end{equation}
where $t$ denotes the shell thickness.  We are interested in soft elastomers that can undergo large deformations, and so we consider the two dimensional compressible Neo-Hookean model for stretching energy \cite{Bonet_1997}:
\begin{equation}
     W_{s} = \frac{E}{4(1+\nu)}\Big (\text{tr}(g) - 2 - \ln(\text{det}(g))\Big ) + \frac{E\nu}{2(1-\nu^{2})}\Big (\text{det}(g)^{\frac{1}{2}} - 1 \Big )^{2}.
    \label{eq:WStretcing}
\end{equation}
The bending energy depends linearly on the curvature through the plane strain moduli:
\begin{equation}
    W_{b} = \frac{E\nu}{2(1-\nu^{2})}tr(\tilde{h})^{2} + \frac{E}{2(1+\nu)} \ tr(\tilde{h}^{2}),
    \label{eq:WBending}
\end{equation}
where the relative curvature is
\begin{equation} \label{eq:relcurv}
    \tilde{h} = h - h^{0},
\end{equation} 
where $h^{0}$ the spontaneous curvature.    

We are interested in photomechanically active materials, where one can induce a spontaneous curvature with illumination \cite{Corbett_2015,Korner_2020}, and therefore we take 
\begin{equation}
    h^{0} = \overline{h} - \kappa(t) ,  \quad \tau \dot \kappa = \alpha \tilde I (m\cdot n) \kappa_0, \quad \kappa(0) = 0,
    \label{eq:ill}
\end{equation}
where $m$ is the light propagation direction, $\kappa_0$ is the unit spontaneous curvature depending on the symmetry of material, $\alpha = c_p G$ is a material and geometry dependent constant where $c_p$ is the photo-compliance,  $G$ is a geometric factor, $\tau$ is the relaxation time, $\tilde I$ is the intensity of illumination, and $\overline{h}$ is the curvature of the structure in the absence of illumination.   In this work, we consider a fixed time, and therefore take $ \kappa =  \alpha \tilde I (m\cdot n) \kappa_0$.  

\subsection{Discrete Formulation} \label{sec:disc}

\paragraph{Kinematics} We follow Weischedel {\cite{Weischedel_2012} }(also {\cite{Grinspun_2003,Chen_2018,Crane_2018}}) to obtain a discrete description of the shell, as shown in Figure \ref{fig:config}.  We discretize the shell with a triangulated mesh associated with nodes $\{ R^a \}$ in the reference shell, and denote the corresponding images $\{ r^a \}$ in the current shell.  The deformation is $\Phi:R^a \to r^a$.  We then exploit the fact that one only needs a local parameterization  to obtain the first and second fundamental form.

Consider a particular element in the reference configuration consisting of three nodes $\{ R^a, R^b, R^c \}$.  The corresponding element in the current configuration has respective nodes $\{r^a, r^b, r^c\}$.  Introduce a local parameterization  where $\xi^1$ goes from $R^a$ to $R^b$ while $\xi^2$ goes from $R^a$ to $R^c$.  We can then obtain a discrete covariant basis in the reference and current configuration as
\begin{equation}
E_1 = R^b - R^a, \quad E_2 = R^c - R^a, \quad e_1 = r^b - r^a, \quad e_2 = r^c - r^a,
\end{equation}
for this particular element.  We can then obtain the discrete contravariant basis as usual.  This enables us to compute the discrete stretch $g$ associated with this particular element using (\ref{eq:gh}).

We now turn to the bending metric.  For a given element, we label the edges according to the vertex opposite the edge. We define an edge normal to be the unit bisector of the normal vectors of the two elements sharing the edge.    Since the triangle joining the mid-points of the edges is congruent with the element (at half the size), we can approximate
\begin{equation}
\frac{\partial n}{\partial \xi^1} = -2 (n^b - n^a), \quad \frac{\partial n}{\partial \xi^2} = - 2 (n^c - n^a).
\end{equation}
We can now compute the material curvature $h$ for this particular element using (\ref{eq:gh}), and the relative curvature using (\ref{eq:relcurv}). Explicit represenations are given in the appendix.

We may also compute the referential area of each element as
\begin{equation}
A = \frac{1}{2} |E_1 \times E_2| =  \frac{1}{2} (\text{det} (E_\alpha \cdot E_\beta))^{1/2} .
\end{equation}

\paragraph{Energy and equilibrium} We can now write the total energy as 
\begin{equation} \label{eq:disc_energy}
    \mathcal{E} = \sum_e  A_e \left(\frac{t}{4}W_{s}(g_e) + \frac{t^{3}}{12}W_{b}(\tilde h_e)\right),
\end{equation}
where $e$ indexes the elements, $A_e$, $g_e$, and $h_e$ are the area, stretch, and relative curvature of element $e$, respectively.

In a typical simulation, we start from the natural (stress-free, illumination-free) reference state, and increment the illumination in small increments to justify our quasi-static treatment.  For each increment, we use Newton-Raphson iteration to minimize the total energy (\ref{eq:disc_energy}) over the nodal positions $\{r^a\}$ to obtain the equilibrium shape of the shell. 

\section{Flat reference: Isotropic spontaneous curvature} \label{sec:flatiso}
\begin{figure}
  \centering
  \includegraphics[width=3in]{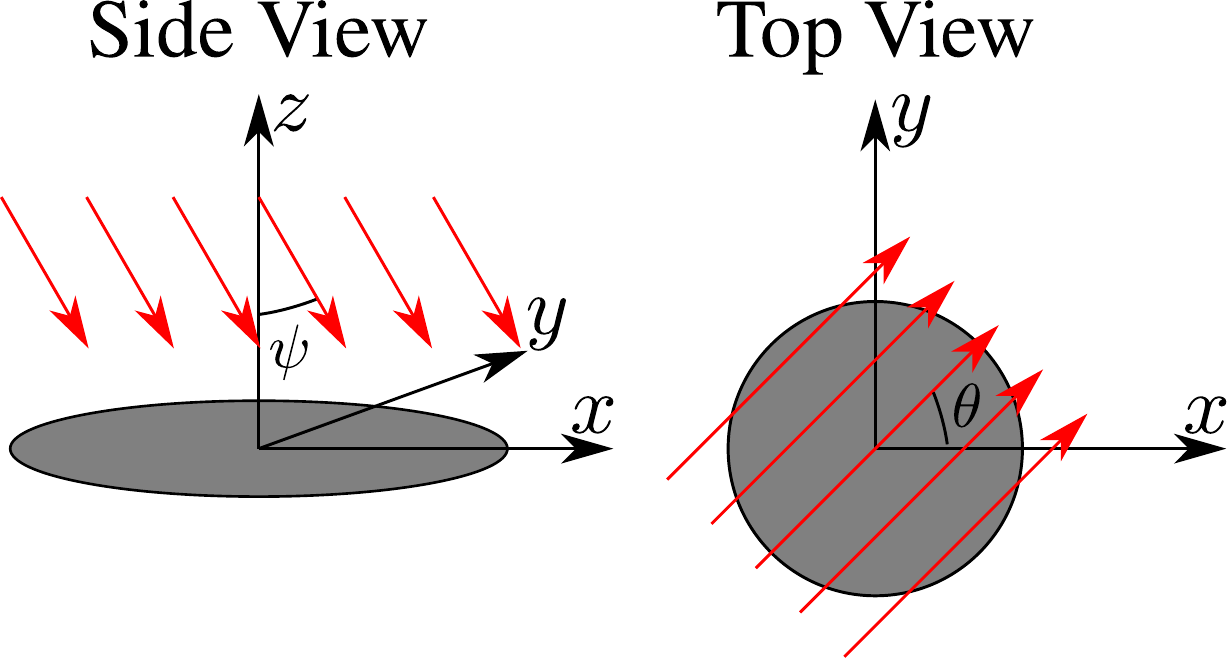}
\caption{The light illuminates a disc set in the $x-y$ plane, with $\psi$ denoting the angle of illumination relative to the vertical $z$ axis, and $\theta$ denoting the projection of the illumination in the $x-y$ in-plane relative to the $x$ axis.}\label{fig:ill}
\end{figure}

\subsection{Flat reference}

We consider a flat thin circular film with diameter 2 cm and thickness 30 $\mu$m unless otherwise specified. We discretize the domain with 12417 nodes, or 37251 degrees of freedom. We illuminate the entire film with a wide circularly polarized parallel beam propagating in a direction
\begin{equation}
m = \{ \sin \psi \cos \theta, \sin \psi \sin \theta, -\cos \psi \},
\label{eq:lightVectr}
\end{equation}
where $\psi$ is the angle from the vertical $z$ axis, and $\theta$ denotes an arbitrary direction in the $x-y$ plane, as shown in Figure \ref{fig:ill}. We non-dimensionalize the length with the film radius $r$, take $\alpha = 0.194$ m/W \cite{Korner_2020}, and non-dimensionalize $I$ with a typical intensity 515 W/m$^2$ such that the common intensity used in this work becomes $I=1$ in non-dimensional units.

We start with a film synthesized in the isotropic state such that $\kappa_0= \text{Id}_2$.

\subsection{Normal illumination}

\begin{figure}[t]
  \centering
  \includegraphics[width=5in]{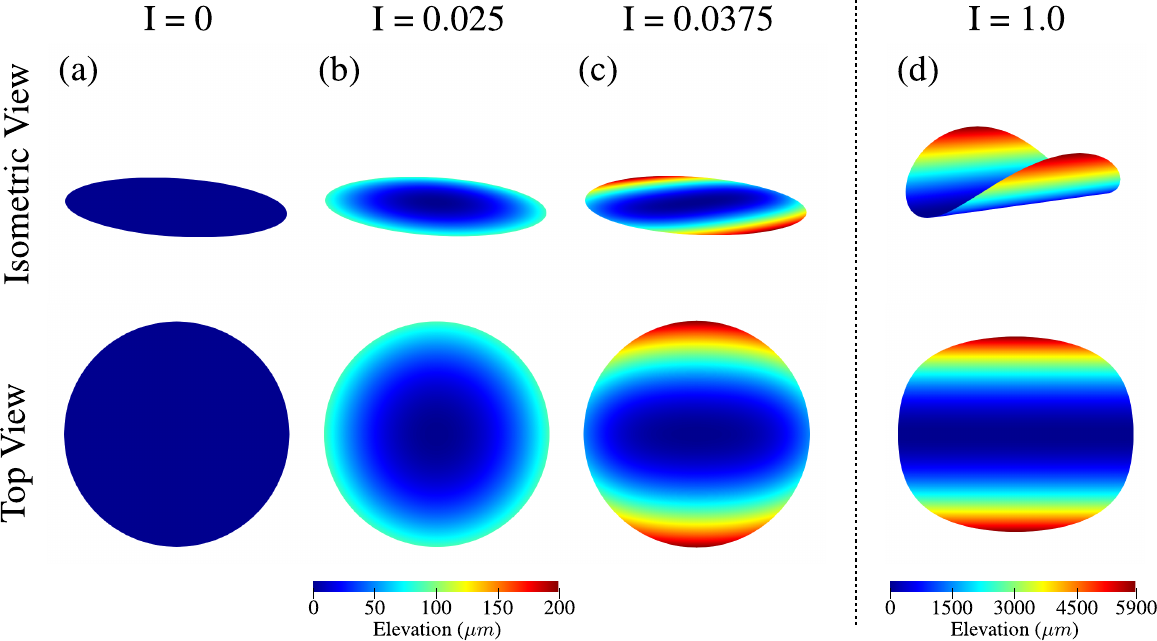}
\caption{A flat isotropic liquid crystal elastomer sheet is illuminated vertically source of light. The reference state at intensity (a) $I=0$ is a flat sheet. When increasing the light intensity to (b) $I = 0.025$, the film edges begin to curve up due to the increase in spontaneous curvature. Further increasing the light intensity to (c) $I = 0.0375$ results in a large buckling deformation, and the film morphs into a cylindrical shell. Further increasing the light intensity to (d) $I = 1.0$ increases the spontaneous curvature, further elevating the edges of this cylindrical shell.}\label{fig:flat_normal}
\end{figure}

We take $m=\{0,0,-1\}$ such that the illumination is normal to the initial flat state.   We start with a (non-dimensional) illumination intensity of $I = 0$ and gradually increase it to $I = 1$.  We fix five nodes at the center of the sheet to simulate a clamped flat constraint at the center. The results are shown in Figure \ref{fig:flat_normal} (See supplementary material for an animation).

When the light is turned on, spontaneous curvature increases, driving the system towards forming a spherical bowl of uniform curvature. However, this deformation induces a change in Gauss curvature and, consequently, stretch (according to Gauss's Theorema Egregium \cite{Gauss_1828}).  Since the stretching energy scales with thickness while bending scales with thickness cubed, the stretching behavior dominates in thin shells.  Therefore the stretching energy suppresses the curvature, leading to a flat sheet with the edges curved up, as shown at $I=0.025$  (Figure \ref{fig:flat_normal}(b)).  The sheet eventually bifurcates into a cylindrical shape, again to minimize stretch, at $I = 0.0375$ (Figure \ref{fig:flat_normal}(c))\footnote{The orientation of the axis of the cylindrical shell is arbitrary, depending on mesh and perturbations in the simulation.}. Post bifurcation, the deformed shell retains this cylindrical shape, and the curvature continues to increase with illumination, shown at $I=1$ (Figure \ref{fig:flat_normal}(d)). This bifurcation to a cylindrical shape is consistent with other works regarding thermally induced spontaneous curvature \cite{Dunn_2007,Hebner_2023}.

\subsubsection{Analytic model}
\begin{figure}
\centering
\includegraphics[width=3in]{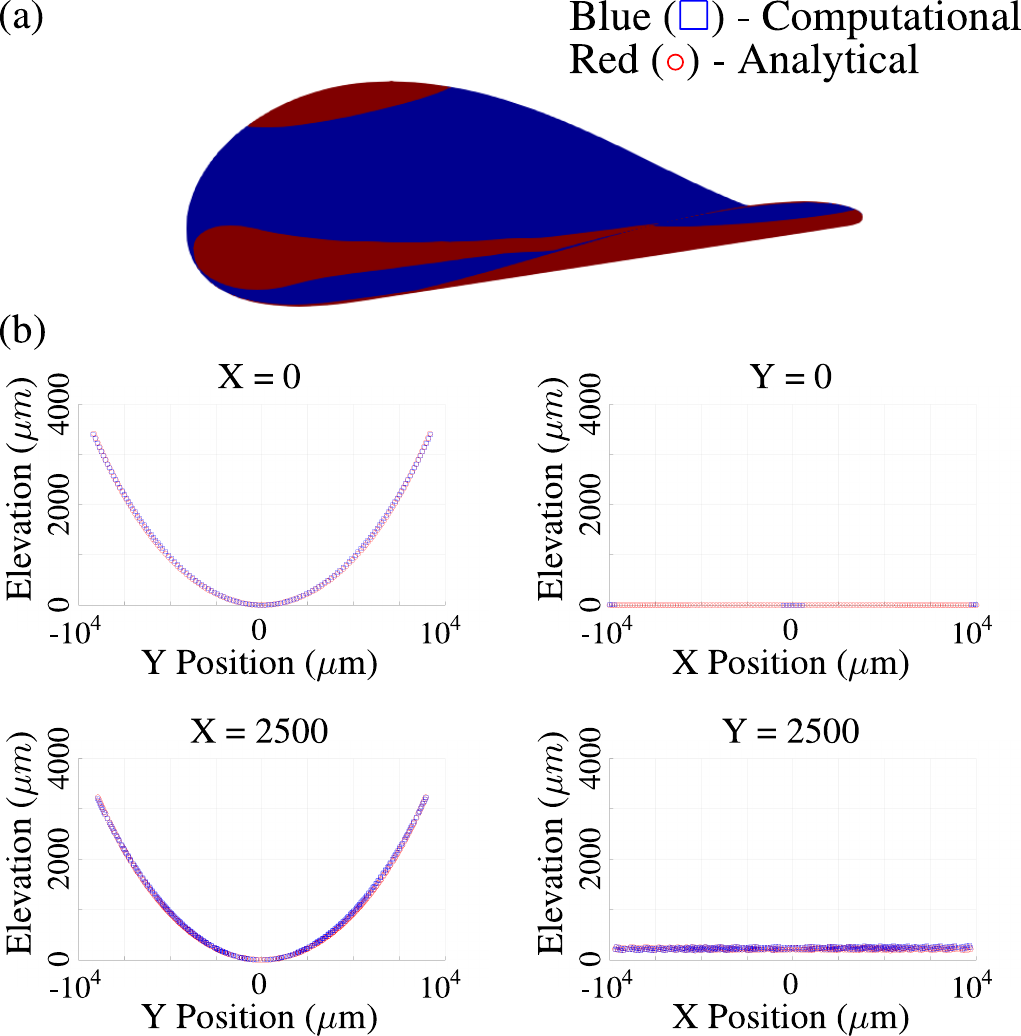}
  \caption{(a) The deformation computed from the discrete shell computational model is shown in blue, overlain with the deformation computed analytically in red. Cross-sections comparing the position of the analytical deformation to the computational result are shown for (b) $\tilde{x} = 0$, $\tilde{x} = 0.25$, $\tilde{y} = 0$, and $\tilde{y} = 0.25$.}\label{fig:analytic}
\end{figure}

We briefly examine the post-buckled shape using a simple analytic model. Informed by the observations made from Figure \ref{fig:flat_normal}, we make the ansatz that the shell deforms to a perfect cylinder with zero stretch, and minimize the energy over the cylinder radius $\rho$ to obtain the deformed shape. 

Let $x,y$ parameterize the disc (such that $\eta^1 = x, \eta^2 = y$ in the notation of Section \ref{sec:cont}), and we assume that the shell deforms such that the axis of the cylinder is parallel to the $x$-axis. We may then respectively write the deformation, curvature, and spontaneous curvature as: 
\begin{equation}
r = \begin{pmatrix} \frac{x}{\rho} \\ \sin \frac{y}{\rho} \\  1- \cos\frac{y}{\rho} \end{pmatrix}, \quad
h =  \frac{1}{\rho} \begin{pmatrix} 0 & 0 \\ 0 & 1 \end{pmatrix}, \quad
h^* = - I  \cos \frac{y}{\rho} \begin{pmatrix} 1 & 0 \\ 0 & 1 \end{pmatrix},
\end{equation}
The total energy consists only of the bending contribution:
\begin{equation}
\mathcal{E} = \frac{Eh^{3}}{12} \int_{-1}^1 \int_{-\sqrt{1-y^{2}}}^{\sqrt{1-y^{2}}} \frac{\nu}{2(1-\nu^{2})}\left(\tr(M)\right)^{2} + \frac{1}{2(1+\nu)} \tr(M^{2}) \ dx dy,
\end{equation}
where 
\begin{equation}
M = h + h^{*} = \begin{pmatrix}  - I  \cos \frac{y}{\rho}  & 0 \\ 0 &  \frac{1}{\rho} - I  \cos \frac{y}{\rho}  \end{pmatrix}
\end{equation}
is the bending matrix. We fix $I = 0.5$ , and minimize this energy over the cylinder radius $\rho$ to obtain the deformed shape. We compare the results of this simplified analytical model to the complete computational model, with the results shown in Figure \ref{fig:analytic}. Figure \ref{fig:analytic}(a) depicts the deformations for illumination intensity $I = 0.5$. We plot various cross-sections of the two deformations in Figure \ref{fig:analytic}(b). The error between the total energy computed for the analytical deformation relative to the computationally-computed deformation is $8.1\%$. Visually, we see excellent agreement between the analytical and computational deformations. Therefore, this simple analytic model is able to accurately capture the deformed shape of the cylinder. 

\subsection{Illumination Direction}

We now break the in-plane isotropy by slanting the direction of illumination from the vertical.  We study two boundary conditions, one where a small region (five nodes) in the center is clamped flat like the previous cases, and a second condition where the sheet is allowed to pivot using an elastic hinge joint (one node fixed and four nodes fitted with elastic springs) attached to the center point.

\subsubsection{Center clamped}
\begin{figure}[t]
  \centering
  \includegraphics[width=5in]{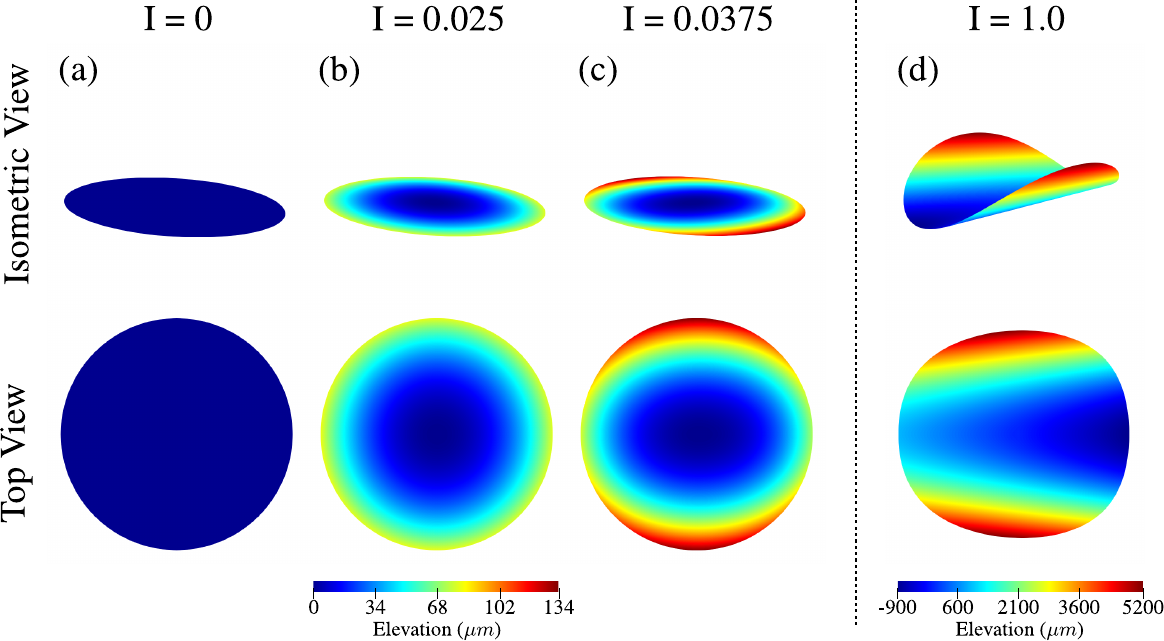}
\caption{A flat isotropic liquid crystal elastomer sheet is illuminated from a parallel source of light angled $\psi = 30^{\circ}$ from the vertical, along the $x$ axis. The reference state at intensity (a) $I=0$ is a flat sheet. Increasing the light intensity to (b) $I = 0.025$, the edges of the sheet begin to curve up due to the increase in spontaneous curvature. Further increasing the light intensity to (c) $I = 0.0375$ results in a large buckling deformation, and the film morphs into a cylindrical shell. Further increasing the light intensity to (d) $I = 1.0$ increases the spontaneous curvature, further elevating the edges of this cylindrical shell.}\label{fig:seq_30_0}
 \end{figure}

We begin with the case where the five nodes nearest to the center are fixed (clamped flat).  We illuminate the sheet at an angle $\psi = 30^\circ$ from the vertical and $\theta = 0^\circ$ in-plane direction, with the resulting deformation shown in Figure \ref{fig:seq_30_0}. We observe a similar sequence as in the case of vertical illumination, with two key differences. 

First, bifurcation occurs at a higher intensity for the angled light ($\psi = 30^{\circ}$) at $I = 0.0375$, when compared to normal illumination ($\psi = 0^{\circ}$) which occurs at $I = 0.0325$. A greater intensity is required because the angled light less directly illuminates the shell, resulting in a lower effective light intensity. Notably, the vertical component of the angled light at bifurcation is identical to the corresponding intensity for vertical illumination within 3 significant figures.

Secondly, the post-bifurcated shape is conical, and the axis of the cone is aligned with the direction of illumination. This conical deformation forms due to the the non-zero angle of incidence of the illumination, which creates an energetically favorable direction for the shell to bifurcate along. The spontaneous curvature is minimized when the illumination direction lies within the tangent space of the shell; bifurcating along the in-plane axis achieves said minimization.

\begin{figure}
  \centering
  \includegraphics[width=5.5in]{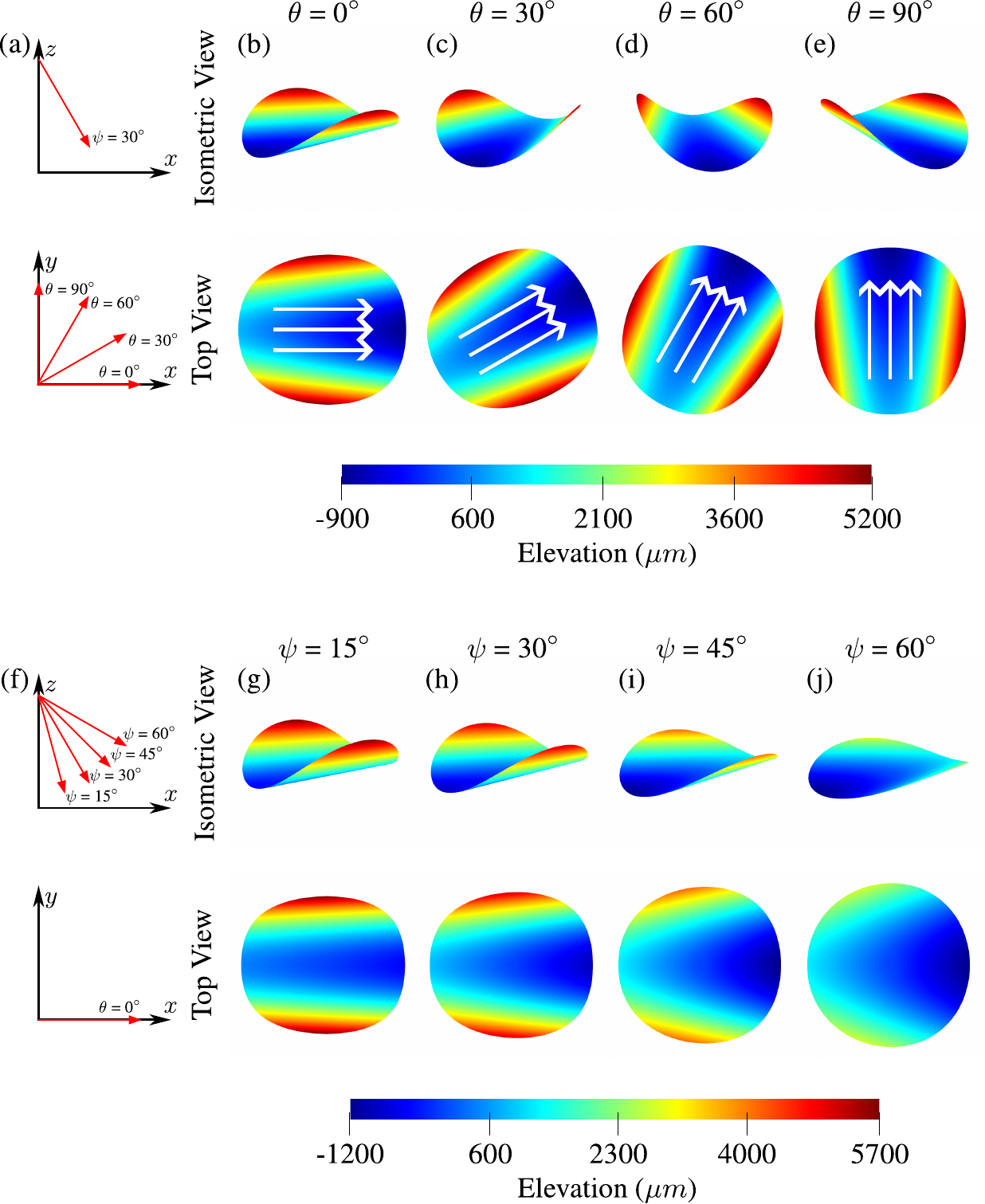}
\caption{A flat sheet is illuminated from the (a) given orientation, corresponding to states (b)-(e). The light intensity and vertical angle are $I = 1.0$ and $\psi = 30^\circ$ respectively. Further, (b)-(e) show an isometric view of the deformed states as well as a top-down view of the corresponding deformations. The in-plane $x-y$ component of the illumination is represented by the white arrows. The deformation aligns along in-plane angle $\theta$ equal to (b) $0^\circ$, (c) $30^\circ$, (d) $60^\circ$, and (e) $90^\circ$. Separately, a flat sheet is illuminated from the (f) given orientation, corresponding to states (g)-(j).  The light intensity and in-plane angle are $I = 1.0$ and $\theta = 0^{\circ}$ respectively. Further, (g)-(j) show an isometric view of the deformed states as well as a top-down view of the corresponding deformations. The deformations are shown for vertical angles $\psi$ equal to  (g) $15^{\circ}$, (h) $30^{\circ}$, (i) $45^{\circ}$, and (j) $60^{\circ}$.}\label{fig:flat_sweep}
\end{figure}

We now compare the effects of varying the direction of illumination, both the in-plane angle $\theta$ and the vertical angle $\psi$. Figure \ref{fig:flat_sweep} shows the post-bifurcated shape for the various illumination directions at $I = 1.0$. Figures \ref{fig:flat_sweep}(a-e) show that changing the in-plane direction $\theta$ controls the orientation of the conical deformation, but not the overall behavior. The spontaneous curvature is minimized when the edges of the conical deformation align parallel to the illumination. In short, the deformed cylinder follows the source of illumination, much like a sunflower tracks the sun's position. An animation of the centrally clamped sheet deforming to follow a moving light source is shown in the supplementary material.

Figures \ref{fig:flat_sweep}(f-j) show the effect of varying the vertical angle of illumination $\psi$. The phenomena remains similar qualitatively: a conical deformation emerges along the direction of the in-plane angle $\theta$. However, as the angle of incidence increases, the magnitude of the deformation decreases. Note that $m \cdot n = \cos \psi$, and therefore, the effective illumination decreases with increasing $\psi$. We compare the critical value of intensity at which bifurcation occurs in Table \ref{tab:crit}. We observe that the critical intensity increases with $\psi$. As the light less directly illuminates the shell, the intensity required to achieve bifurcation increases.

\begin{table}
\centering
\begin{tabular}{ |c|c|c|c|c|c| }
\hline
Case & $\psi = 0^\circ$ & $\psi = 15^\circ$ & $\psi = 30^\circ$ & $\psi = 45^\circ$ & $\psi = 60^\circ$ \\
\hline
Flat &0.0325  & 0.0375 &0.0375 &0.0450& 0.0625 \\
\hline
Spherical &0.0575 &0.0625&0.0650 &0.0775&0.1100\\
\hline
\end{tabular}
\caption{Critical intensity for bifurcation}\label{tab:crit}
\end{table}

\subsubsection{Pinned: The artificial sunflower}
\begin{figure}
  \centering
    \includegraphics[width=4in]{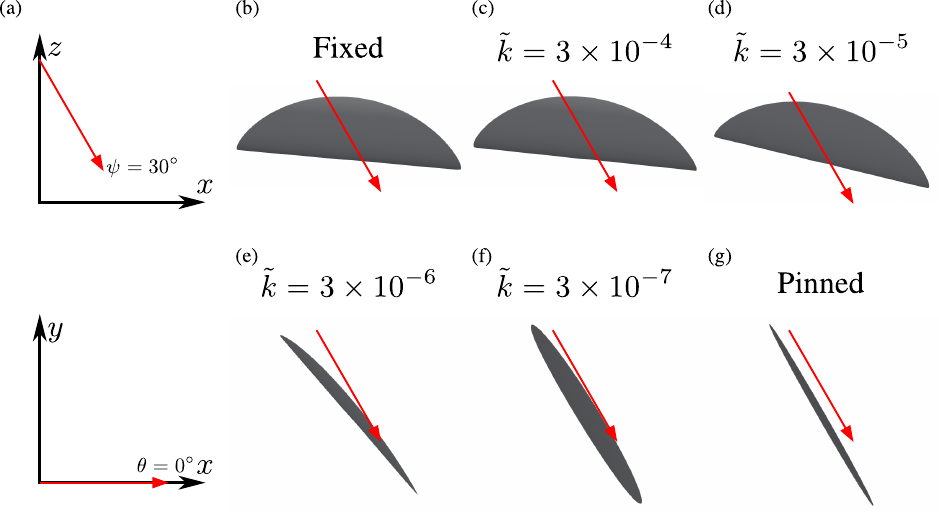}
  \caption{An artificial sunflower: a sheet with a pin joint at the center and torsional spring. The deformation is shown for (a) intensity $I = 1.0$, $\psi = 30^{\circ}$ and $\theta = 0^{\circ}$. We model a (b) fixed center, (c) $\tilde{k} = 3 \times 10^{-4}$, (d) $\tilde{k} = 3 \times 10^{-5}$, (e) $\tilde{k} = 3 \times 10^{-6}$, (f) $\tilde{k} = 3 \times 10^{-7}$, and (g) a pinned center. The direction of illumination is depicted by a red arrow.}\label{fig:SpringLight}
\end{figure}
We now alter the system constraints: we fix a single center point and attach springs to the four surrounding nodes to imitate a torsional spring. We modify the energy as
\begin{equation}
    \mathcal{\tilde{E}}^{new} = \mathcal{\tilde{E}} + \frac{1}{2}\tilde{k}\overset{4}{\underset{i = 1}{\sum}}\vert r^{i} - R^{i}\vert^{2},
\end{equation}
where $\mathcal{\tilde{E}}^{new}$ is the new energy to minimize, $\tilde{\mathcal{E}}$ is the non-dimensional energy, $\tilde{k}$ is the non-dimensional spring constant, and vertices $i$ are the four nodes adjacent to the center. A torsional spring allows for rotation with an energy penalty dependent on the spring constant. We investigate this deformation for a fixed angle of incidence $\psi = 30^{\circ}$ (Figure \ref{fig:SpringLight}(a)). We model a fully clamped center (Figure \ref{fig:SpringLight}(b)), torsional springs of varied strength (Figure \ref{fig:SpringLight}(c-f)), and a pinned center (Figure \ref{fig:SpringLight}(g)).

As the stiffness of the simulated elastic hinge decreases, the shell rotates further towards the light and undergoes a smaller bending deformation. For a fully pinned shell, there is zero deformation, just pure rotation. This behavior can be predicted from (\ref{eq:ill}), as the change in spontaneous curvature scales with $m\cdot n$, and pure rotation can satisfy equilibrium with zero spontaneous curvature. Thus, these solutions reflect that for a specific value of spring constant, the sheet trades bending energy for spring energy; changing the spring constant can alter this balance.

We can exploit this behavior for various applications, such as fabricating self-aligning structures. Instead of expensive sensors and actuators, space structures can contain azobenzene-doped liquid crystal elastomers to function as alignment devices. Additionally, spacecrafts require temperature regulation, and one common method is using radiators to reject excess heat towards dead space. Incorporating liquid crystal elastomers can allow for the replacement of expensive, manually-controlled actuation systems with passive alignment mechanisms, enabling radiators to automatically adjust their alignment for optimal heat rejection.

\section{Flat reference: Patterned director} \label{sec:flataniso}

Various researchers have developed techniques to synthesize nematic liquid crystal elastomer films with a specified director orientation in the natural reference state \cite{McConney_2013,Ware_2015,Gelebart_2017_2, Kowalski_2017}. Such films have been extensively studied for thermal actuation, where altering the temperature leads to spontaneous stretch \cite{Biggins_2012}]  However in this work, we hold the temperature constant, resulting in zero spontaneous stretch. Instead, we examine systems where illumination generates anisotropic spontaneous curvature determined by the written director orientation. In particular, we consider unit spontaneous curvature
\begin{equation}
    \kappa_0 = \hat{d} \otimes \hat{d} - \frac{1}{2} \hat{d}^{\perp} \otimes \hat{d}^{\perp},
\end{equation}
where $\hat{d}$ is the local orientation of the liquid crystal director, and $\hat{d}^{\perp}$ is the perpendicular in-plane direction with respect to the director. Consequently, the film is anisotropic. 
Additionally, the director does not have be be uniform across the film. 
We model a similar problem as before: shining light at a liquid crystal elastomer sheet clamped at the center.

\subsection{Circular Sheet with Azimuthal Defect}
\begin{figure}
  \centering
  \includegraphics[width=2.5in]{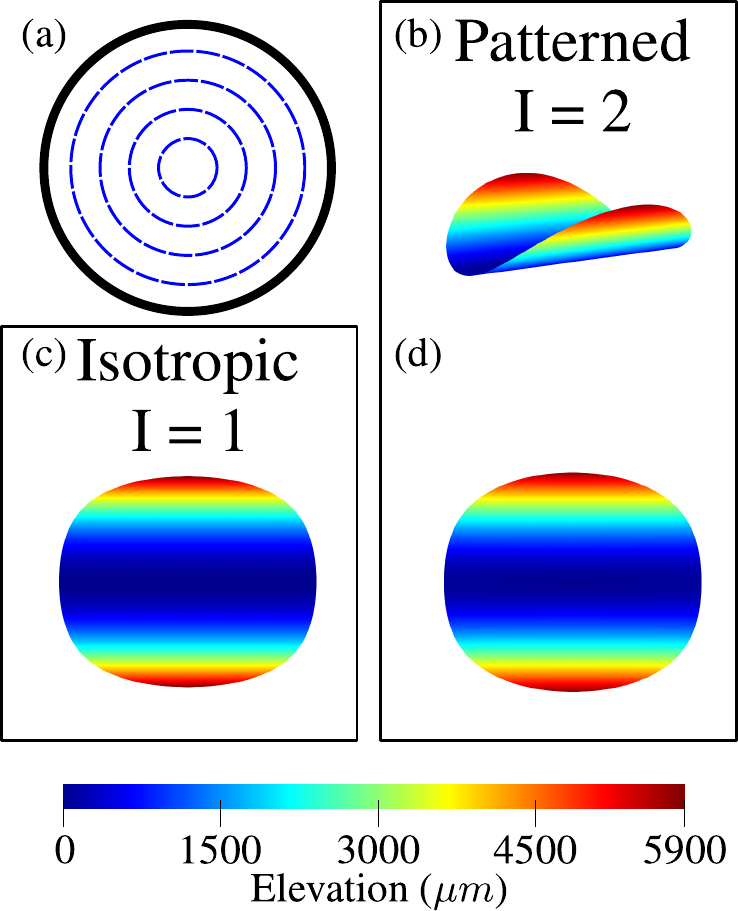}
  \caption{(a) We program a +1 azimuthal defect into a flat disc. The illumination is directly vertical with $\psi = 0^{\circ}$. The patterned sheet is then illuminated to $I = 2.0$ (b)(d). We compare the deformation of the patterned sheet to the deformation of the isotropic film at lower intensity $I = 1$.}\label{fig:disclination}
\end{figure}

We consider a liquid crystal elastomer film with a director field having a +1 azimuthal defect, illustrated in Figure \ref{fig:disclination}(a). The pattern and the illumination are axisymmetric in-plane, resulting in an initially axisymmetric deformation. Similar to the isotropic disc, the film bifurcates into a cylindrical shell at a critical intensity, with the final deformation shown in Figures \ref{fig:disclination}(b,d) for intensity $I=2$.  However, the curvature is smaller than that of the isotropic sheet: Figure \ref{fig:disclination}(c) shows the result for an isotropic sheet at lower intensity $I=1$. The unit spontaneous curvature $\kappa_{0}$ in the patterned film causes positive bending along one direction and negative curvature in the perpendicular direction. This difference in the bending directions reduces the effective spontaneous curvature relative to the isotropic film.

\subsection{Rectangular Sheet with Constant Director Orientation}
\begin{figure}
  \centering
 \includegraphics[width=4in]{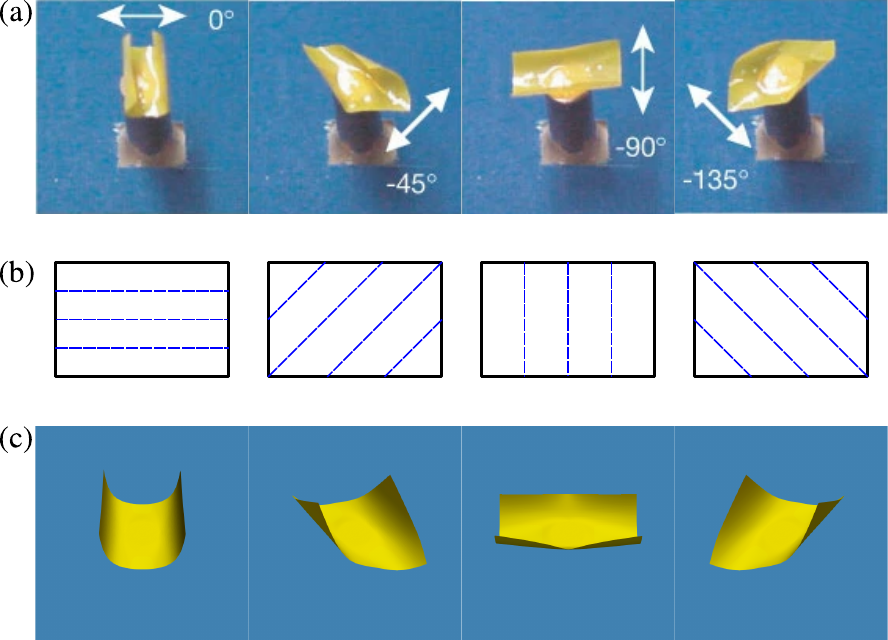}
  \caption{The (a) experiments from Yu \textit{et al.} \cite{Yu_2003} are shown for various linear polarizations of light. Rather than using linearly polarized light, circularly polarized light is used to illuminate various liquid crystal elastomer sheets with (b) constant patterned director orientations. The (c) resulting deformations are shown for $I = 5.0$.}\label{fig:LightDirectorSquare}
\end{figure}

We now turn to the seminal experiments of Yu \textit{et al.} \cite{Yu_2003}, who used linearly polarized light to illuminate a rectangular sheet, with Figure \ref{fig:LightDirectorSquare}(a) adapted from their work. We consider various uniform director patterns over rectangular liquid crystal films, depicted in Figure \ref{fig:LightDirectorSquare}(b). The uniform alignment of the director field imitates the effect of linearly polarized light illuminating an isotropic film. By varying the alignment of the director field by 45$^{\circ}$ and using a circularly polarized light source, we can replicate the effects of varying the linear polarization of a light source over an isotropic film.

We consider a rectangular sheet $2$ $cm$ by $1.3$ $cm$ with a $0.5$ $cm$ diameter circular region clamped flat. We discretize the domain with 10965 nodes, or 32895 degrees of freedom. The simulation results for the given director orientations are shown in Figure \ref{fig:LightDirectorSquare}(c). 
The film bends perpendicular to the director orientation, due to the contraction along that direction. We observe good qualitative agreement between these simulations and the experimental observations.

\section{Shells} \label{sec:shell}

We now examine the deformation of curved shells, which have an inherent curvature prior to illumination. Therefore, the spontaneous curvature will have a contribution from both the photo-responsive behavior and the reference geometry of the shell.

\subsection{Spherical Cap}
We consider the deformation of a spherical cap, under similar conditions as previous analysis: the shell is an isotropic liquid crystal and the center is clamped flat. The spherical cap has diameter $2$ $cm$ and thickness $30$ $\mu m$, and has radius of curvature $10.025$ cm. We discretize the domain with 12417 nodes, or 37251 degrees of freedom.

\begin{figure}
  \centering
  \includegraphics[width=4in]{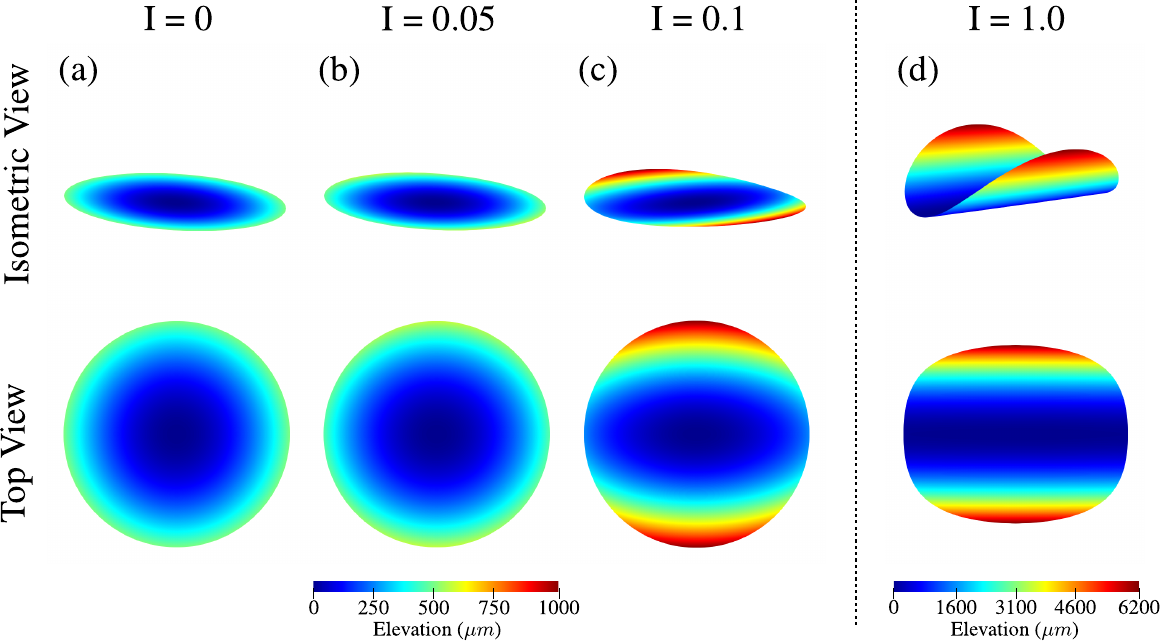}
  \caption{A curved spherical shell made from an isotropic liquid crystal elastomer is illuminated from a directly vertical source of light. The reference state at intensity (a) $I=0$ is a flat sheet. Increasing the light intensity to (b) $I = 0.05$, the edges of the sheet begin to curve up due to the increase in spontaneous curvature. Further increasing the light intensity to (c) $I = 0.1$ results in a large buckling deformation and the spherical cap morphs into a cylindrical shell. Further increasing the light intensity to (d) $I = 1.0$ increases the spontaneous curvature, further elevating the edges of this cylindrical shell.
  }\label{fig:CapStudy}
\end{figure}

We  illuminate the spherical cap vertically, starting with illumination intensity $I = 0$ and gradually increasing to $I = 1.0$. The results are shown in Figure \ref{fig:CapStudy}. The initial behavior is qualitatively similar to the normal illumination of a flat sheet (Figure \ref{fig:flat_normal}): the edges of the shell rise up in the shape of a cap, and upon some reaching some critical intensity, the shell bifurcates into a cylindrical shape. However, the intensity required for bifurcation is greater for the spherical cap when compared to the flat sheet, as shown in Table \ref{tab:crit}. This increased critical intensity arises from the inherent reference curvature of the spherical cap, which provides additional rigidity \cite{Reddy_2006}.  The behavior as we vary the angle of illumination is also similar as in the case of a sheet (see Figure \ref{fig:CapPsiTheta} in the supplementary material). An animation of the cap being subjected to a moving illumination source is shown in the supplementary material. 

\subsection{Cylindrical Shells}

\begin{figure}
  \centering
\includegraphics[width=5in]{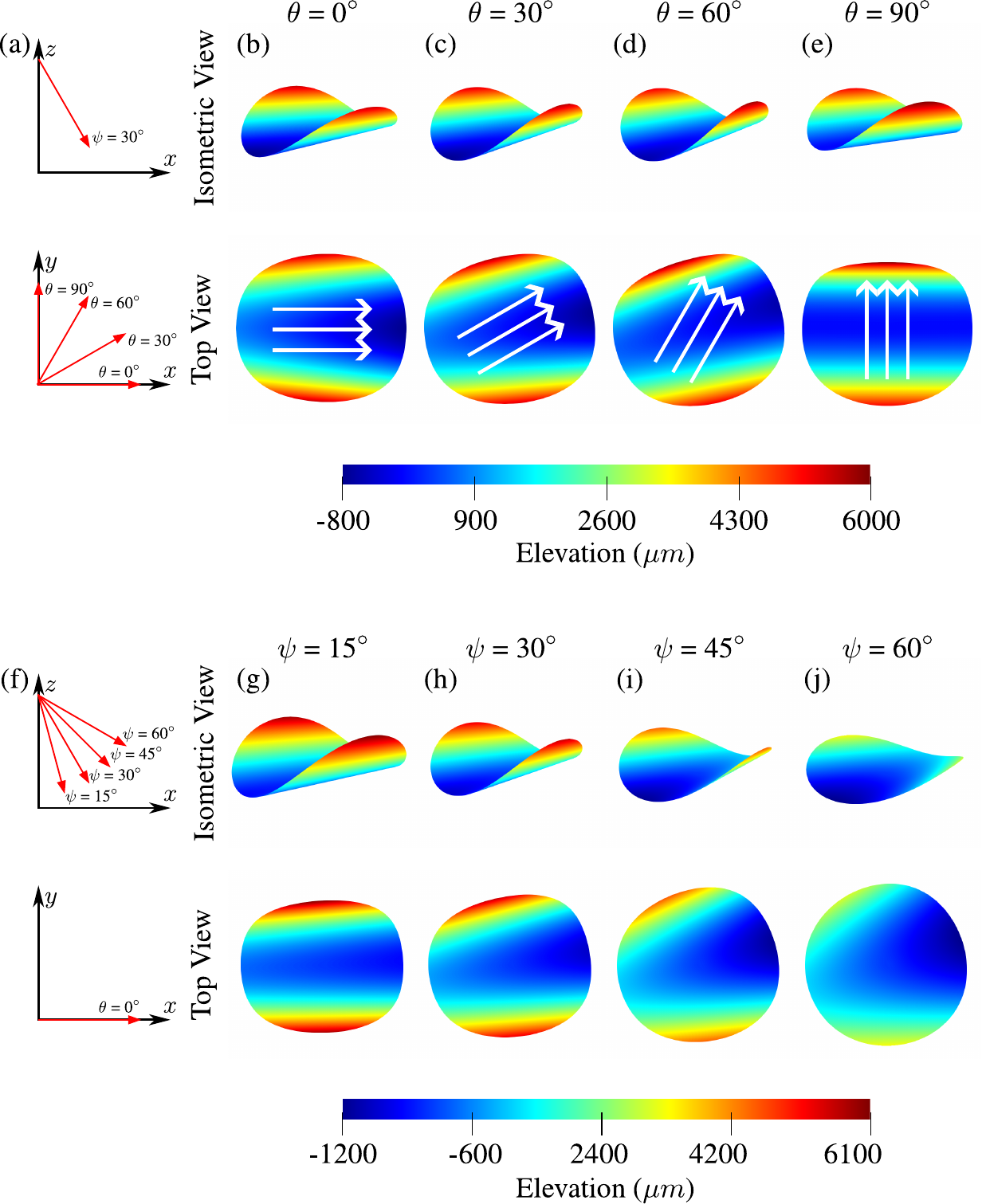}
\caption{A cylindrical shell is illuminated from the (a) given orientation, corresponding to states (b)-(e). Further, (b)-(e) show an isometric view of the deformed states as well as a top-down view of the corresponding deformations. The in-plane $x-y$ component of the illumination is represented by the white arrows. The deformation aligns along the (b) $0^\circ$, (c) $30^\circ$, (d) $60^\circ$, and (e) $90^\circ$ in-plane angle $\theta$. The light intensity and vertical angle are $I = 1.0$ and $\psi = 30^\circ$ respectively. A cylindrical shell is illuminated from the (f) given orientation, corresponding to states (g)-(j). Further, (g)-(j) show an isometric view of the deformed states as well as a top-down view of the corresponding deformations. The vertical angle from the $z$ axis is given for (g) $\psi = 15^{\circ}$, (h) $\psi = 30^{\circ}$, (i) $\psi = 45^{\circ}$, and (j) $\psi = 60^{\circ}$. The light intensity and in-plane angle are $I = 1.0$ and $\theta = 0^{\circ}$ respectively.}
\label{fig:CylinderPsiThetaSweep}
\end{figure}

We now turn to a shell with a non-axisymmetric reference configuration.  In the preceding examples of axisymmetric shells, the symmetry was broken by the direction of illumination.  In non-axisymmetric shells we anticipate competition between the reference shape and the illumination direction.  To explore this behavior, we examine cylindrical shells with radius of curvature $10.025$ $cm$, initially aligned along the $x$ axis.  We then illuminate the structure from various directions, with the results shown in Figure \ref{fig:CylinderPsiThetaSweep}.

We hold the vertical angle $\psi = 30^{\circ}$, and vary the in-plane angle $\theta$ relative to the axis of the reference cylinder, with the results shown in Figures \ref{fig:CylinderPsiThetaSweep}(b-e).  When the illumination aligns with the axis of the reference cylinder (Figures \ref{fig:CylinderPsiThetaSweep}(b)), we see an increase in curvature (the edges rise higher), and that cylinder morphs into a cone.  The axis of the resulting cone is aligned with both the illumination and reference cylinder.  As we change the (horizontal) angle of illumination (Figures \ref{fig:CylinderPsiThetaSweep}(c,d)), the deformed shape is still conical, but the axis of the cone lies in between the axis of the reference cylinder and direction of illumination.  The axis is restored when the illumination is perpendicular to cylinder axis, and the shape is that of a distorted cylinder. The animation of the perpendicular case is provided in the supplementary information.


%
%

\section{Conclusions} \label{sec:conc}
Stimuli-responsive materials have garnered significant attention for their potential applications in soft robotics and actuators. Among this class of material, liquid crystal elastomers are noteworthy due to their ability to react to various stimuli. In this work, we developed a discrete shell model to explore the effects of illumination on photoactive liquid crystal elastomer shells.   We use a discrete shell kinematics that enables the stretching and bending to be computed solely from nodal positions, and therefore provides computational simplicity and efficiency compared to traditionally shell models that require higher order discretization.

We utilized this model to examine the deformations of various films and shells under illuminated.  For an axisymmetric structure with a clamped center, large bending deformation arise from illumination, causing the shell to bifurcate into a cylindrical or conical shape. Furthermore, these deformations align with the direction of light. If the clamped center is replaced with a universal joint, the shell can rotate to better align with the illumination direction; the degree of bending deformation relative to the degree rotational motion can be tuned, allowing strong control over the energy dissipation mechanism. For liquid crystalline sheets in the nematic state, director orientation can be a substitute for linearly polarized light to achieve directional bending deformations.
For a non-axisymmetric structures, like a cylindrical shell, the photo-responsive behavior and reference geometry compute to influence the deformation alignment.

With these modeling capabilities, we can explore a greater scope of problems. We have already adapted this model for thermal stimulation in stimuli-responsive materials, and have observed great agreement with experimental observations. Moreover, we can apply this discrete shell model to rapidly simulate passive materials, which lends itself to being a powerful tool in optimization algorithms. This model helps capture behavior of liquid crystal elastomers, and enhancing these design capabilities can provide greater engineering freedom when designing actuators and self-aligning structures.


\section*{Acknowledgment}
We gratefully acknowledge \cite{Evouga_Github} which we used for the meshing and calculation of first and second fundamental forms.  This work has been funded by the US Office of Naval Research (MURI grant N00014-18-1-2624).

\appendix
\section{Explicit formulas}
The discrete first fundamental form $a$ can be calculated over a triangle with vertices $\{v_{a},v_{b},v_{c}\}$ as:
\begin{equation}
    a = \begin{bmatrix}
        (v_{c} - v_{a}) \cdot (v_{c} - v_{a}) & (v_{b} - v_{a}) \cdot (v_{c} - v_{a}) \\
        (v_{c} - v_{a}) \cdot (v_{b} - v_{a}) & (v_{b} - v_{a}) \cdot (v_{b} - v_{a})
    \end{bmatrix}.
\end{equation}
The discrete first fundamental form $b$ can be calculated as:
\begin{equation}
    b = 2\begin{bmatrix}
        (n_{c} - n_{a}) \cdot (v_{c} - v_{a}) & (n_{b} - n_{a}) \cdot (v_{c} - v_{a}) \\
        (n_{c} - n_{a}) \cdot (v_{b} - v_{a}) & (n_{b} - n_{a}) \cdot (v_{b} - v_{a})
    \end{bmatrix}.
\end{equation}

\section{Illumination Intensity Calculation}
We aim to compute $\alpha$, which is a material and geometry dependent constant which relates illumination intensity to spontaneous curvature. We take:
\begin{itemize}
    \item Photocompliance: $c_{p} = 5.4 \times 10^{-5} \frac{m^{2}}{W}$ \cite{Korner_2020},
    \item Penetration Depth: $d = 0.56 \mu m$ \cite{Korner_2020},
    \item Thickness: $t = 30 \mu m$ (Film Dimension),
\end{itemize}
and compute the effective macroscopic coupling constant
\begin{equation}
    \alpha = -\frac{12 c_{p}}{t^{3}}\int_{-\frac{t}{2}}^{\frac{t}{2}}exp\left ( -\frac{\frac{t}{2}-z}{d}\right)zdz
\end{equation}
to be
\begin{equation}
    \alpha = 0.194 \frac{m}{W}.
\end{equation}
When non-dimensionalizing (\ref{eq:relcurv}), we can compute
\begin{equation}
    \overline{I} = \frac{1}{\overline{x}\alpha} = \frac{1}{(0.01m)(0.194 \frac{m}{W})} = 515\frac{W}{m^2},
\end{equation}
where $\overline{I}$ corresponds to non-dimensional intensity $I = 1.0$ throughout the rest of this paper.

\newpage
\bibliographystyle{abbrv}
\bibliography{bibliography.bib}

\newpage
\renewcommand{\thepage}{S\arabic{page}} 
\setcounter{page}{1}
\renewcommand{\thefigure}{S\arabic{figure}}
\setcounter{figure}{0}
\begin{center}
\section*{Supplementary Information}
\noindent
Artificial sunflower: Light induced deformation of photoactive shells\\
S. Sanagala and K. Bhattacharya
\end{center}

\begin{figure}
  \centering
  \includegraphics[width=4in]{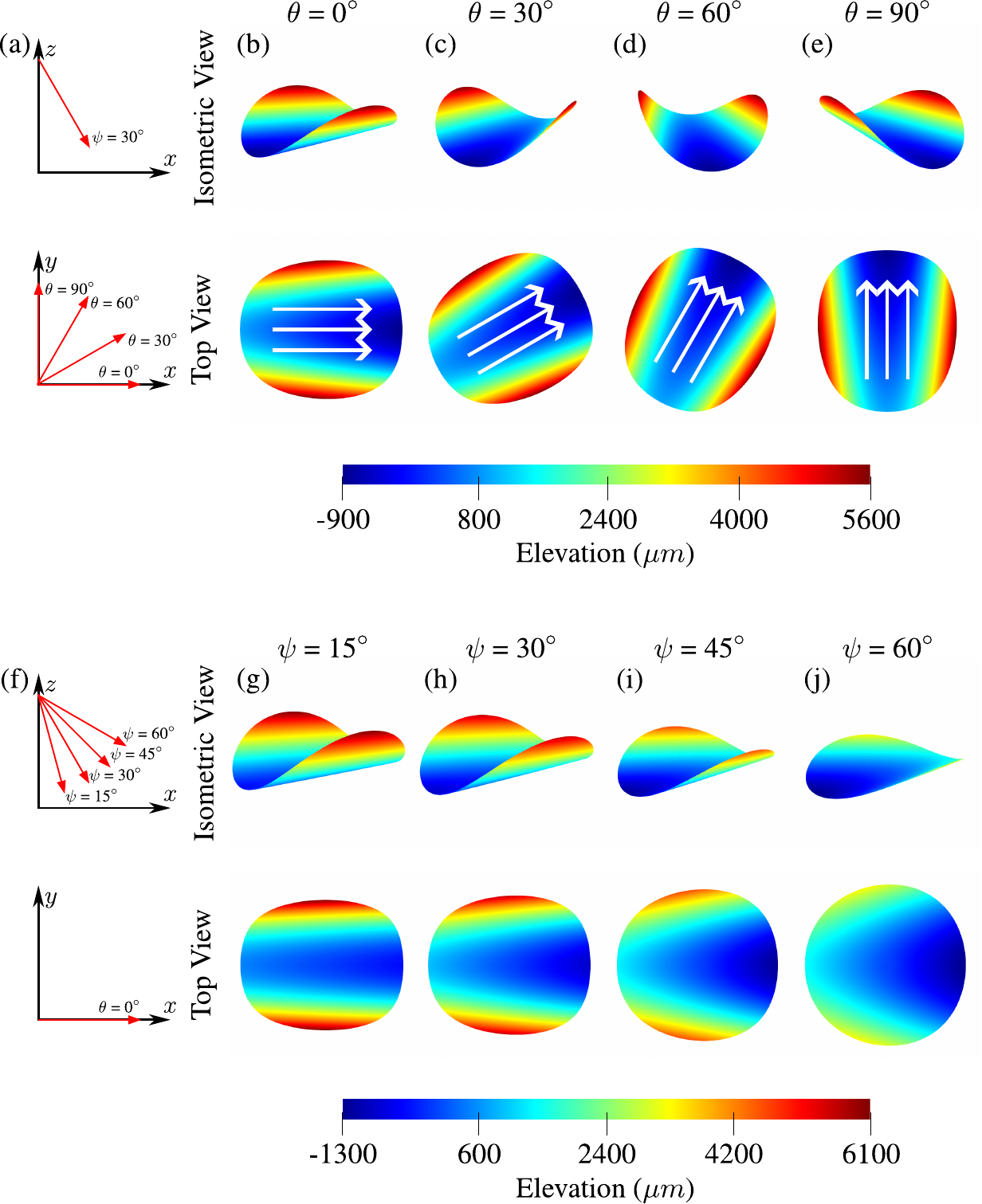}
  \caption{A spherical shell is illuminated from the (a) given orientation, corresponding to states (b)-(e). Further, (b)-(e) show an isometric view of the deformed states as well as a top-down view of the corresponding deformations. The in-plane $x-y$ component of the illumination is represented by the white arrows. The deformation aligns along the (b) $0^\circ$, (c) $30^\circ$, (d) $60^\circ$, and (e) $90^\circ$ in-plane angle $\theta$. The light intensity and vertical angle are $I = 1.0$ and $\psi = 30^\circ$ respectively. A spherical shell is illuminated from the (f) given orientation, corresponding to states (g)-(j). Further, (g)-(j) show an isometric view of the deformed states as well as a top-down view of the corresponding deformations. The vertical angle from the $z$ axis is given for (g) $\psi = 15^{\circ}$, (h) $\psi = 30^{\circ}$, (i) $\psi = 45^{\circ}$, and (j) $\psi = 60^{\circ}$. The light intensity and in-plane angle are $I = 1.0$ and $\theta = 0^{\circ}$ respectively. }\label{fig:CapPsiTheta}
\end{figure}

\paragraph{List of Figures}
\begin{enumerate}
\item Figure \ref{fig:CapPsiTheta}: A spherical shell illuminated from various angles.
\end{enumerate}

\paragraph{List of Movies}
\begin{enumerate}
\item S1\_Normal\_Illumination.mp4: This video depicts normal illumination of a flat sheet, as the edges rise up and the sheet bifurcates. This corresponds to Figure \ref{fig:flat_normal}.
\item S2\_Flat\_Clamped\_Swivel.mp4: This video depicts a clamped flat sheet being illuminated, initially incrementing from $I = 0$ to $I = 1$, with $\theta = 0^{\circ}$ and $\psi = 30^{\circ}$. Next, the angle $\theta$ increases with $I$ and $\psi$ held constant, until a full rotation is made. Then $I$ is subsequently decreased to $I = 0$.
\item S3\_Flat\_Pin\_Swivel.mp4: This video depicts a pinned flat sheet being illuminated, initially incrementing from $I = 0$ to $I = 1$, with $\theta = 0^{\circ}$ and $\psi = 60^{\circ}$. Next, the angle $\theta$ increases with $I$ and $\psi$ held constant, until a full rotation is made. Then $I$ is subsequently decreased to $I = 0$. The sheet rotates to remain parallel to the illumination direction.
\item S4\_Cap\_Swivel.mp4: This video depicts a spherical cap being illuminated, initially incrementing from $I = 0$ to $I = 1$, with $\theta = 0^{\circ}$ and $\psi = 30^{\circ}$. Next, the angle $\theta$ increases with $I$ and $\psi$ held constant, until a full rotation is made. Then $I$ is subsequently decreased to $I = 0$.
\item S5\_Cylinder\_Swivel.mp4: This video depicts a cylindrical shell being illuminated, initially incrementing from $I = 0$ to $I = 1$, with $\theta = 0^{\circ}$ and $\psi = 30^{\circ}$. Next, the angle $\theta$ increases with $I$ and $\psi$ held constant, until a full rotation is made. Then $I$ is subsequently decreased to $I = 0$.
\end{enumerate}

\end{document}